\documentclass[10pt]{article}
\usepackage{graphicx,amsmath,amssymb}
\begin{document}
\title{Structure evolution of Pd-Ta-H alloy  In Edwards'
thermodynamics representation}
\author{V.M. Avdyukhina,  A.A. Katsnelson\footnote{E-mail: albert@solst.phys.msu.su},
A.I. Olemskoi, \\ D.A. Olemskoi, G.P. Revkevich, E.A. Goron \\
\textit{Moscow State University, Physics Department }\\
\textit{119899, Moscow, Russia}}
\date{}
\maketitle

\begin{abstract}
X-rays diffraction pictures time dependence of deformed alloy
Pd-Ta being charged with hydrogen has been shown to be possibly
caused by multi-pits character of energetic relief in the states
space. Phenomenological model representing alloy structural
evolution as a occasional roaming on minima of internal energy of
non-ergodic system, has been offered in Lorenz synergetic scheme
frames. Here, order parameters are the part of minima occupied by
the system, conjugated field is considered to be Edwards entropy
and control parameter is taken to be internal energy.
Thermodynamics interpretation of Pd-Ta-H alloy evolution structure
as a complex non-ergodic system is offered
\end{abstract}

This work is supported with RFBR grant (99-02-16135).

\section{Introduction}

The main object of statistics physics has lately become complex
non-ergodic systems - spin and structural glasses, non-ordered
heteropolymers, granulated matter, transportation flow, etc (see
\cite{1},). The main peculiarity of such systems is phase space
separated into isolated regions with every region representing
meta-stable thermodynamic state and their number $N_0$
exponentially exceed full number of quasi particles $N \rightarrow
\infty$, N: $N_0 = \exp \left( sN \right)$ , where s is so-called
Edwards entropy respective to every complexity \cite{2,3}.  Unlike
Bolsman's measure describing disorder in given statistics
ensemble, Edwards entropy describes disorder in complex system
states distribution on internal energy minima respective to
statistics ensemble. Statistics ensemble's play the role of
particles in complicated systems.   The distribution on these
ensembles is described with effective temperature T and entropy S,
introduced by Edwards \cite{3}.  So, complicated system with $T=0$
is corresponding to granulated matter distributed eventually on
all minima  (flat distribution).

Edwards systems that have exponentially huge number of minima
$N\gg 1$,  give us an example which is  opposite to ergodic
systems where $N_ 0 = 1$.  According to this, an intermediate case
where the number of minima $N_0$   is neither exponentially big
nor equal to one is of great interest.

Deformed Pd-Ta alloys being electrolytically charged with hydrogen
represents this example as it is shown below.   Really, solid
state material under the influence of strong deformation can go so
far away from the equilibrium state that the self-organization
effects acting in dissipative structure formation become essential
\cite{4,5}. During hydrogen charging of metal its atoms create
essential internal tensions that can lead the system into a
non-equilibrium state as an external influence does. The
researches of  annealed Pd-W-H  alloy \cite{6} è deformed Pd-Sm-H,
Pd-Er-H, Pd-Mo-H and Pd-Ta-H alloys \cite{7,8,9,10,11,12,13} show
that the non-monotonous structure transformation of non-regular
character can take place there. Using of synergetic models
\cite{6,10} allows to explain the main features of these systems
behavior. But these models give no explanation of peculiarities of
Pd-Ta-H alloy structural evolution shown below (see also
\cite{12}).   This paper is devoted to this problem solution.

\section{ Experimental data}

The method of the investigations was described in details in
\cite{9,10,11,12}, so we point out only at the fact that the
original charging of Pd-Ta deformed sample ($7 at.\%Ta$) has been
done electrolytically for 15 minutes at current density of 40
mA/cm2. After relaxing for 176 hours, the sample has been charged
for the second time for the same period of time at 80 $ mA/ñm ^2
$. X-Rays researches of diffraction  maxima 220 è 311 were done
with computerized diffractometer with using  $Cu K_{\alpha 1}$
components of x-rays spectrum doublet.  The diffraction maxima
decomposition into components was completed with "Origin" program
package taking into consideration Lorenz form of components
\cite{14,15}.

The led researches show that, after the hydrogen charging of Pd-Ta
deformed alloy, regular displacement of diffraction maxima which
represents the lattice getting bigger and after that smaller take
place. During the relaxation, stochastic changes of diffraction
maxima position, their broadness and symmetry appear. So does
stochastic transformation of their form including appearance and
disappearance of some picks.   Diffraction curves and their
decompositions received for 200 maxima after the secondary
charging at the different moments of time are shown at the fig.1.
They represent complex structure of diffraction picks and
non-monotonous character of their changes with time. Actually, the
diffraction maxima have the structure shown at the table here.

As a bell-like form of diffraction maxima reflects homogeneous
phase, one can presume that such changes of diffraction pictures
are consequences of phases mutual transformations accordingly to
Pd, Ta and H atoms re-distribution over different regions of the
system. These structural changes can be repeating but not
periodic.  On the other hand, different broadness of components
and their non-monotonous time changes mean that this process is
accompanied with non- monotonous defect structure evolution and
elastic fields created by this evolution appearance. Along with
essential difference between hydrogen binding energy with ideal
and defect Palladium lattice \cite{16}, the existence of
far-acting fields causes essentially non-equilibrium character of
researched system \cite{5}.  As it is known from the theory of
glass-like theory \cite{1}, it causes complicated energetic relief
appearance in the phase space of non-equilibrium state.  Character
of this relief is defined by the original defect structure and
alloy atoms distribution, that are formed by mechanical work up,
regimes of hydrogen charging and following relaxation.   The
difference of time dependence for the 220 and 331 coherent
scattering regions oriented differently to the surface of the
sample points out to energy being passed from freedom degrees to
others that can lead to diffusion flows turbulence \cite{12}.

\section{ Theoretical scheme}

The non-monotonous structural changes discussed above can be
explained with the preposition that Pd-Ta deformed alloy hydrogen
charging leads to the set of meta-stable states appearance
corresponding to different phases and defect structures.  In this
case, generalized system diffusion over internal energy minima in
accordance with  the pointed states leads to stochastic structural
changes observed during the experiment.

To understand this phenomenon, one needs to take into
consideration the non-monotonous character of evolution, which is
not of periodic character.   These changes remind of strange
attractor behavior described with synergetic Lorenz scheme
\cite{17}. Earlier, using Lorenz scheme has allowed us to describe
non-monotonous behavior of Pd-Er-H double-phased alloy \cite{6}.
While doing that, we parameterized the system with value portion
of phase enriched with Er, concentration of Er atoms there and Er
atoms traps concentration in matrix.

This parameterization cannot be used in multi-phased Pd-Ta-H alloy
where the situation is much more difficult.  Facing this problem
for the first time, using parameters generalized over all the
phases or yielding the most typical one helps avoiding the
necessity of many phases behavior description.    But this way we
could have lost the extremely important peculiarity of
multi-phased Pd-Ta-H alloy evolution:  not only value correlation
of the phases but their number change with time.  According to
this, the problem of parameters choice for the synergetic model
which tries  to describe Pd-Ta-H system stochastic behavior
becomes very important.

According with Ruelle-Takens theorem \cite{18}, non-trivial
behavior of such system is observed if the number of
parameterizing freedom degrees is over two.   At phase transition,
the system behavior is described with hydrodynamic mode and its
amplitude represents order parameter defined with thermostat
state. Self-organized (synergetic) system peculiarity is: not only
 thermostat's influence over the subsystem but also its
influence over the thermostat is very important. This influence
can be either direct or indirect. First of them is defined with
conjugated field, the second one - with control parameter. So, for
the alloys of ordered synergetic model , order parameter is in
accordance with usual long range order parameter, conjugated field
refers to difference between chemical potential of components and
control parameter - to difference of unit cell knots population of
different atoms \cite{19} Usual dissipative regime of phase
transition is realized if relaxation time for order parameter and
conjugated field is much longer then it is  for control parameter.
Otherwise, stationery value of control parameter increasing over
crucial value leads not to ordering of this system but to
transforming into regime of strange attractor \cite {20}.

To explain non-monotonous change of Pd-Ta-H alloy we take into
consideration the fact that hydrogen charging leads to internal
energy complicated relief state formation in the space.  This
relief has many minima separated with barriers (multi-valley
structure). This preposition usage explains critical slowing down
of structural transformation in the Pd-H systems \cite{21}.  This
slowing down is caused by the hierarchical portrait of the relief;
big minima are overlapped with smaller ones, then the smaller
minima are covered with the minima that are even smaller, etc.
Because of that, during its evolution, system is enforced to fill
the smallest minima first, starting to fill bigger and bigger
minima up to the biggest one which  describes  the system as a
whole subject.

As to the observed case, the consideration which is limited with a
type of minima that are not essentially different from each other
is enough for our purpose.  We numerate them with index $\alpha$
covering the set of numbers 1, 2,.. $N_0$. As a result, the system
evolution is described with probability of distribution over these
minima $p_{\alpha}$ changes. Defining of
$p_{\alpha}\left(t\right)$ time dependence requires Fokker-Planck
equation to be solved which is very difficult \cite{22}. However,
for the preliminary stage, we consider the integral
characteristics that describe the type of $p_\alpha$ distribution
and thermodynamical behavior of the system as a whole subject
enough for investigation.  We take  parameters as following:

\begin{itemize}
\item
Number N of internal energy minima where the system is situated at
the moment (obviously, it describes a half of dispersion width for
probability $p_\alpha$ );

\item
Entropy of the system dispersion over those minima
\begin{equation}
 S=-\sum_{\alpha}{p_{\alpha}\ln{p_{\alpha}}}   \label{1}
\end{equation}

\item
Specific internal energy $e_\alpha$ at given minimum $\alpha$,
which defines the full value of the internal energy
\begin{equation}
e=-\sum_{\alpha}{ p_{\alpha} e_{\alpha} }   \label{2}
\end{equation}
\end{itemize}

We will use  the relation $n = N/N_0$ of number N of  internal
energy minima occupied by system and full number of minima in the
system $N_0$ as an order parameter.     Then specific Edwards
entropy $s = S/N_0$ which defines disorder in the distribution
over minima  $\alpha$ is described with  conjugated field and
internal energy  density and acts as a control parameter.

For the phenomenological description of the system evolution, main
parameters $\frac{dn}{dt}$, $\frac{ds}{dt}$, $\frac{de}{dt}$ of
 velocity changes  are to relate to their values n, s, e. Here, Lorenz
system advantage is in its accordance to the simplest choice of
Hamiltonian in the respective microscopic representation \cite{5}.
Linear Lorenz equation is
\begin{equation}
 \frac{dn}{dt}=-\frac{n}{\tau_n}+g_{n}s,  \label{3}
\end{equation}

where first part describes Debuy's relaxation of  n(t) value to  n
= 0 with character time $\tau_n$; second part which contains
positive coefficient $g_n$  describes number  of internal energy
minima occupied by the system at increasing of entropy s
 in the dispersion over those minima.

Unlike (\ref{3}), equations describing velocities $\frac{ds}{dt}$,
$\frac{de}{dt}$ contain non-linear parts that represent the
influence of the subsystem over the thermostat, showing the
feedback we have discussed before.  Thus, the equation for entropy
change is
\begin{equation}
 \frac{ds}{dt}=-\frac{s}{\tau_s}+g_{s}ne,  \label{4}
\end{equation}

where first part describes Debuy's relaxation to  s = 0 with time
$\tau_{s}$.   In accordance to second principle of thermodynamics,
non-linear part which containes positive coefficient $g_s$ takes
into consideration entropy s increasing because of system
dispersion over less deep minima of internal energy.

Last of the required equations defines velocity of internal energy
increasing.
\begin{equation}
 \frac{de}{dt}=-\frac{e-e_{0}}{\tau_e}+g_{e}sn  \label{5}
\end{equation}

Here, Debay's relaxation with character time $\tau_e$  leading not
to zero value of the internal energy e but to $e_0$   defined by
system position at the phase diagram and the alloy preliminary
work up is taken into consideration.  Non-linear part with
positive coefficient $g_e$ represents negative feedback meaning
that the re-distribution over internal energy minima should lead
to decreasing of its full value e because of the system
transformation to deeper minima. It should be pointed out to a
very important role of non-linear parts.  Their competition
describes self-organized system behavior: positive feedback in
(\ref{4}) causes entropy s increasing because of relation between
relative number of minima n and internal energy e. Positive
feedback in equation (\ref{5}) describes internal energy e
decreasing because of the correlation between the number of minima
n and entropy s.

The system of differential equation (\ref{3})-(\ref{5}) generally
has no analytic solution.  Using of non-dimensional values n, s,
e, t scaling them as shown below seems to be very suitable:

\begin{equation} \label{6}
n_{m}=
{\left({\tau_{s}g_{s}}\right)}^\frac{1}{2}{\left({\tau_{e}g_{e}}
\right)}^\frac{1}{2}, \qquad
s_{m}=\frac{n_{m}}{\tau_{n}g_{n}},\qquad
 e_{m}= \left(
\tau_{n}g_{n} \right)^{-1} \left( \tau_{s}g_{s}
\right)^{-1},\qquad \tau_{n}.
\end{equation}

As a result, the equations (\ref{3})-(\ref{5}) appear to be

\begin{equation}
 \frac{dn}{dt}=-n+s,  \label{7}
\end{equation}

\begin{equation}
 \tau\frac{ds}{dt}=-s+ne,  \label{8}
\end{equation}

\begin{equation}
 \Theta\frac{de}{dt}=\left(E-e\right)-sn,   \label{9}
\end{equation}
where the  non-dimensional parameters are entered
\begin{equation}
 \tau=\frac{\tau_{s}}{\tau_{n}}, \qquad
\Theta=\frac{\tau_{e}}{\tau_{n}},  \qquad  E=\frac{e_{0}}{e_{m}}.
\label{10}
\end{equation}

In adiabatic regime $\tau$, $\Theta \ll 1$  the principle of
hierarchic co-ordering    takes place.  In accordance with this
principle, changes of conjugated field s(t) and control parameter
e(t) follow order parameter $n(t)$ changes \cite{18}. In this
case, left parts of the equations (\ref{8}), (\ref{9}) can be
omitted and they lead to:
\begin{equation}
 s=\frac{En}{1+n^{2}}, \qquad
 e=\frac{E}{1+n^{2}}.
 \label{11}
\end{equation}

In accordance with thermodynamics principles, they mean that
Edwards' entropy monotonously grows but internal energy falls down
with phases number growing up.  On the other hand, variable n
excluding from the equalities  (\ref{11}) gives simple dependence
of entropy on internal energy:
\begin{equation}
s=\sqrt{e\left(E-e\right)}. \label{12}
\end{equation}

Using the definition of the temperature

\begin{equation}
 T\equiv\frac{\partial e}{\partial s}.
 \label{13}
\end{equation}

we get
\begin{equation}
 T=-{\left(1-\frac{E}{2e}\right)}^{-1}\sqrt{\frac{E}{e}-1}.
 \label{14}
\end{equation}

It means that Edwards' temperature becomes negative in the
interval of the internal energy values $E/2< e < E$ and the system
is in the process of self-organization.  Therefore, in accordance
with definition  (\ref{13}) at $T < 0$ every increasing of
internal energy $\delta e > 0$ leads to entropy decreasing
$\delta s < 0 $, i.e. to self-ordering.

Substitution of the first of the equalities (\ref{11}) into
equation (\ref{7}) turns Lorenz' system into Landau-Khalatnikov
equation.
\begin{equation}
 \frac{\partial n}{\partial t}=-\frac{\partial W}{n},
 \label{15}
\end{equation}

where the role of the free energy is played by synergetic
potential

\begin{equation}
 W=\frac{1}{2}n^{2}-\frac{E}{2}\ln{\left(1+n^{2}\right)}
 \label{16}
\end{equation}

At little values of internal energy E which is saved by the system
as a result of  external influence, dependence W(n) monotonously
increases with minima at n = 0 and multi-phased state is not
advantageous synergetically. However, exceeding over crucial value
E = 1 synergetic potential W(n)  reaches minima at
\begin{equation}
 n_{0}=\sqrt{\left(E-1\right)}
 \label{17}
\end{equation}
                                                  .

Therefore, dissipative process of the system relaxation into the
state respective to the value $n_{0}\not =0$  of relative number
of the phases appears to be advantageous. During this process,
Edwards' temperature (\ref{14}) reaches stationary value
\begin{equation}
T_{0}=-\frac{\sqrt {\left( E-1 \right)} }{1-E/2}, \label{18}
\end{equation}

and , being negative at over-critical values $E > 1,$ monotonously
decreases with growing of the internal energy saved by the system
in consequence of external influence.

The represented adiabatic regime $\tau,$ $\theta\ll 1$ reflects
monotonous transition of the dissipative system into non-ergodic
stationary  state. As we are interested in non-monotonous
behavior, we can make a conclusion that the character time
$\tau,\Theta$ ratio should not be small.   If it happens,
analytical investigation of Lorenz' system appears to be
impossible but the analysis \cite{20} show that the regime of
strange attractor of the self-organizing system specific for the
experimental situation is realized at the condition $\Theta
>\tau > 1.$  Thus, time of internal energy change should exceed
respective values of entropy and number of minima occupied by the
system. Obviously, it is realized in the experiment investigated
\cite{12}.

Besides these limitations for the ratio of relaxation time,
strange attractor existence requires execution of $E > 1$
condition \cite{20}.  This means that the stochastic changes of
the structure in consequences of system states re-distribution
over the minima respectively to different phases are realized only
at the condition that the value of the internal energy saved by
the system in a result  of external influence exceeds critical
value $e_m $ which is defined with the third expression of
(\ref{6}). It can be noticed, that the systems that have great
value of positive feedback's linear constant $g_n$    and
non-linear constant $g_s$ and long relaxation time $\tau_n,$
$\tau_s$  are predisposed to stochastic behavior.

\section{Conclusion}

The led analyses shows that the stochastic structure
transformation is realized because of the competitive influence of
order parameter over the conjugated field and control parameter of
self-organizing system.  In the case of multi-phased Pd-Ta-H alloy
 where the roles of two last parameters are
played by specific value of entropy s and internal energy e this
competition is supported with the basic thermodynamic equality:
\begin{equation}
f=e-Ts
\label{19}
\end{equation}

where  T -  Edwards' effective temperature.  This equality is a
result of Edwards' temperature definition (\ref{13}) and
conjugated entropy definition.
\begin{equation}
s=-\frac{\partial f}{\partial T}, \label{20}.
\end{equation}

One of us has shown recently \cite{23} that the pointed
definitions along with identity (\ref{19}) - are originally from
the simplest field  scheme in the limitation of which
self-organizing system behavior is parameterized with
double-component fields of order parameter, conjugated field and
control parameter.   Here first components are reduced to the
values used before-respective number of occupied minima n, entropy
s and internal energy e of non-ergodic system.  Second components
represent generalized flows conjugated to the pointed values such
as flow of the probability density $\vec q$ re-distribution among
minima of internal energy; thermodynamics force $-\nabla f$ equal
to gradient of free energy, taken with the opposite sign;
temperature opposite gradient $-\nabla T.$ These components change
appears to be essential, if the system is at non-stationary state.
Its example represents non-monotonous behavior of multi-phased
Pd-Ta-H alloy. Hopefully, the synergetic scheme offered here
represents the variant of strongly non-equilibrium state systems
thermodynamics theory and its development is far from being
completed  \cite {24,25}.

According to the analysis shown above, kinematics condition of the
system of non-monotonous behavior consists of internal energy
changes and temperature gradient changes that exceed respective
value of entropy and gradient of free energy as well as of minima
occupied by the system and conjugated probability flow. This
condition explains abnormally big values of time intervals where
stochastic behavior appear.   Dynamic condition requires entropy
increasing to have essential influence over the increasing number
of minima where the system is situated . On the other hand, it
requires decreasing of internal energy as consequences of
re-distribution in order to lead to essential increasing of the
entropy itself. Obviously, these conditions are provided with the
essential values of $g_n$ and $g_s$ parameters that are realized
at not high barriers separating minima and low "rigidity" of the
internal energy relief.  Thus, thermodynamics condition is
considered a requirement of strongly non-equilibrium state of the
system.  This state of the system is reached with preliminary work
up of the system such as radiation, essential deformation,
hardening, etc. For the Pd-Ta-H system being investigated this
work up consisted of preliminary deformation and hydrogen charging
that lead to elastic tensions that are about $10\%$ of elastic
module character value. Considering this reasoning, we can presume
that the internal energy saved by the system is big, too.

Let's point out at two observations of general character. First
one is- critical slowing down of the structural transformation
found in the system Pd-H \cite{21} can be also found in the
Pd-Ta-H alloy considered here. Our preliminary data received after
hydrogenation for several times confirms this presumption.  This
means that at the quantity of hydrogenations essential increasing,
the considered above multi-valley structure which is represented
with the same minima transforms into multi-leveled hierarchical
structure of energetic relief, which leads to the system evolution
slowing down.  This slowing down was found by the authors of
\cite{21}.

Second observation is about the data of multi-phased deformed
Pd-Mo-H alloy where the non-monotonous structural changes that are
analog to the ones considered above were found \cite{26}. A s for
Pd-Ta-H regular displacement of diffraction maxima takes place.
This process reflects the lattice parameters getting bigger and,
after that, smaller in anisotropic way.  During the following
relaxation, stochastic changes of the diffraction maxima
components number, their positions, their width and intensity
occur.   Using this data, one can conclude that the non-regular
behavior of the strongly dissipative multi-phased system is a
specific feature for the structural changes of the alloys after
hydrogen charging.

\newpage

\begin{figure}
  \centering
  \includegraphics{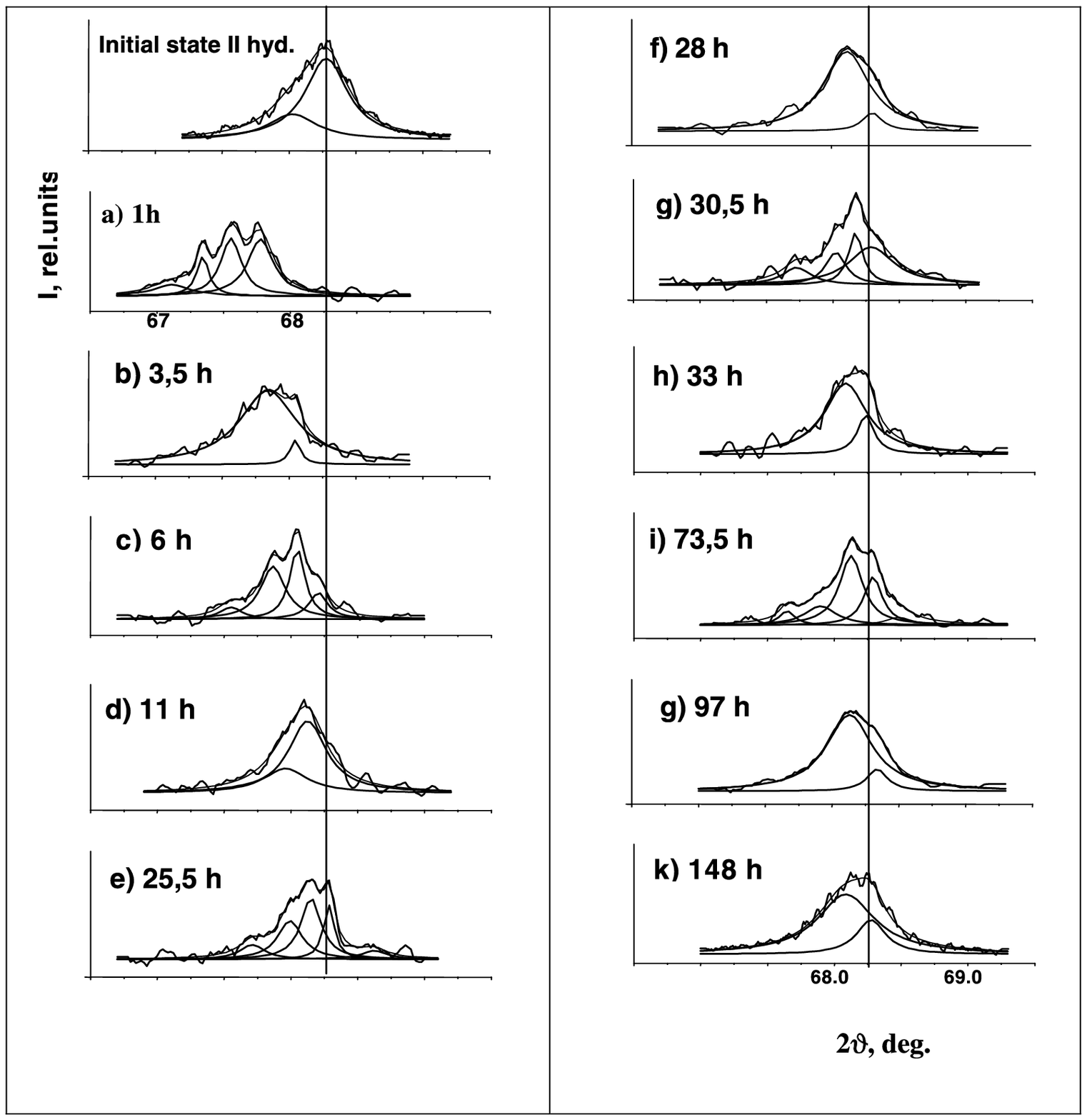}
  \caption{Time change of the position and profile of the diffraction
line (220) after second charging: a) 1 h;  b) 3,5;  c)  6; d) 11;
e) 25,5; f) 28;  g) 30,5; h) 33; i) 73,5;  j) 97,5; k) 148
(vertical line shows the position respective to original condition
- 173 h of relaxation after second charging) }\label{fig}
\end{figure}

\begin{table}  \label{table}
\begin{tabular}{|c|c|c|c|c|c|c|c|c|c|c|c|}
\hline
  \centering
  Time  after the charging, h
  & 1,0 & 3,5 & 6,0 & 11,0 & 25.5 & 28,0 & 30,5 & 33,0 & 73,5 & 97,0 & 148 \\
  Number of picks, N
  & 4 & 2 & 4 & 2 & 5 & 1 & 4 & 2 & 4 & 2 & 2 \\
  Relative number of picks, n
  & 0,8 & 0,4 & 0,8 & 0,4 & 1 & 0,2 & 0,8 & 0,4 & 0,8 & 0,4 & 0,4 \\
  Number of figure & a & b & c & d & e & f & g & h & i & j & k \\ \hline
\end{tabular}
  \caption{Time dependence of diffraction maxima (220) number}
\end{table}

\end{document}